\documentclass[superscriptaddress,twocolumn,nofootinbib,pre,showpacs,preprintnumbers]{revtex4-1}
\usepackage{amsmath,amssymb}
\usepackage{epsfig,graphics,color,calc,graphicx}
\usepackage{hyperref}
\usepackage{mathrsfs}
\usepackage{bigints}

\newcommand{\bb}[1]{{\mathbb{#1}}}

\begin{document}

\title{WIMP detection and slow ion dynamics in  carbon nanotube arrays}

\author{G. Cavoto}
\email{gianluca.cavoto@roma1.infn.it}
\affiliation{INFN Sezione di Roma, Piazzale A. Moro 2, I-00185 Roma, Italy.}

\author{E.N.M.\ Cirillo}
\email{emilio.cirillo@duniroma1.it}
\affiliation{Dipartimento SBAI, Sapienza Universit\`a di Roma, Via A.\ Scarpa 16, I--00161, Roma, Italy.}
 
\author{F. Cocina}
\affiliation{Dipartimento di Fisica, Sapienza Universit\`a di Roma, Piazzale A. Moro 2, I-00185 Roma, Italy.}

\author{J. Ferretti}
\affiliation{Dipartimento di Fisica and INFN, Sapienza Universit\`a di Roma, Piazzale A. Moro 2, I-00185 Roma, Italy.}

\author{A.D. Polosa}
\email{antonio.polosa@roma1.infn.it}
\affiliation{Dipartimento di Fisica and INFN, Sapienza Universit\`a di Roma, Piazzale A. Moro 2, I-00185 Roma, Italy and CERN, Theory Division, Geneva 1211, Switzerland.}

\begin{abstract}
Large arrays of aligned carbon nanotubes (CNTs), open at one end, could be used as target material for the directional detection of weakly interacting dark matter particles (WIMPs).
As a result of a WIMP elastic scattering on a CNT, a carbon ion might be injected in the body of the array and  propagate through multiple collisions within the lattice.
The ion may eventually emerge from the surface with open end CNTs,  provided that its longitudinal momentum is large enough to compensate energy losses and its transverse 
momentum approaches the channeling conditions in a single CNT. Therefore, the angle formed between the WIMP wind apparent orientation and the direction of parallel carbon nanotube 
axes must be properly chosen. We focus on  very low ion recoil  kinetic energies, related to low mass WIMPs ($\approx 11$~GeV) where most of the existing experiments have low sensitivity.
Relying on some exact results on two-dimensional lattices of circular obstacles, we study the low energy ion motion in the  transverse plane
with respect to CNT directions.  New constraints are obtained on how to devise the CNT arrays to maximize the target channeling efficiency.
\end{abstract}


\maketitle

\section{Introduction}
\label{intro}
Directional detection is one of the research frontiers on WIMPs and several techniques are being studied and developed. 
In a recent paper~\cite{Capparelli:2014lua}, it has been proposed that large arrays of aligned carbon nanotubes (CNTs) might be of use as target material for the directional detection of Dark Matter (DM) Weakly Interacting Massive Particles (WIMPs).

As a result of a WIMP scattering on the surface of a CNT, a carbon ion, with energy up to few tens keV, might be extracted\footnote{In this paper we assume totally ionized carbon nuclei.} and propagate through multiple collisions within the CNT empty volume, provided that its initial velocity is sufficiently aligned with the CNT axis: a sort of  {\it  ion channeling} phenomenon.  
However, in a large aligned CNT array, the ion might   be equally channeled  by the interstices among different nanotubes, as mentioned in~\cite{Capparelli:2014lua}. 
The purpose of this paper is to  illustrate the dynamics in these regions and, by means of Monte Carlo (MC) simulations, to provide an evaluation of the ``channeling efficiency" of a large number of CNT arrays as the target of a directional DM detector\footnote{CNT arrays, consisting of $10^{12}-10^{14}$ carbon nanotubes each, are the building blocks of our target. A large number of them must be packed, to obtain a target surface of the order of 100 m$^2$.}. Here, by channeling efficiency we mean the capacity of the target to drive the struck ions towards the top end of the system, with kinetic energy above a certain threshold.
\footnote{For more details, see Sec. \ref{Monte Carlo estimates of the CNT array efficiency}.}

At first, we consider particles satisfying channeling conditions. A continuum repelling potential, from the collective structure of atoms on the CNT surface, can be used to describe the channeling kinematics~\cite{Lindhard-channeling,Artru}.
For nanotube radii much larger than the typical C electron orbital size, the potential can be well approximated by a step  barrier. In this approximation, CNTs are considered as full solid cylinders, totally reflecting those ions having the right initial  conditions, namely a transverse energy lower than the continuum potential barrier \cite[Fig. 4]{Capparelli:2014lua}. The ion motion in the transverse plane is that of a particle bouncing in a 2D lattice of circular obstacles. Longitudinal energy and velocity along CNTs' axes are conserved. Ions are assumed to move freely between consecutive reflections and their trajectories are a combination of broken lines~\cite{Dedkov}. 
Given these initial assumptions, we can validate our Monte Carlo (MC) simulations on known results on classical billiards~\cite{MZ,Cristadoro}.

In a second step, the initial conditions of the extracted ions are computed from the recoiling ion momentum distribution \cite{Capparelli:2014lua}.
When  channeling conditions are not met, the scattered ion is allowed to penetrate the CNT and it  is then subject to larger energy losses \cite{Dedkov,ZBL,O'C-B,Nastasi} and angular deviation effects \cite{Meyer:1971,Sigmund:1974}, due to the interactions with 
atoms at the CNT surfaces.

The results obtained here give more strength to  the proposal presented in~\cite{Capparelli:2014lua} and give important suggestions on how to devise the CNT arrays. 
Although we are still in the stage of a conceptual design and theoretical considerations, the outcome of this work is another step towards the testing of the prototype envisaged in~\cite{Capparelli:2014lua}.

\section{Directional dark matter detection}
With really small expected rates, the challenge of dark matter detection experiments is to distinguish an unambiguous signal from a large background. 

Direct detection experiments identify the  recoils produced by incident WIMPs on the detector target nuclei. The signal is subject to an annual modulation, related to the variation of the Earth's relative motion with respect to the DM Galactic halo.
Up to now, only the DAMA experiment \cite{Bernabei:2015xba} claimed  the detection of an annual modulation with an undeniable statistical significance. CRESST \cite{CRESST} and CoGeNT \cite{CoGeNT} also reported anomalous results, but their consistency with DAMA ones is a vexed question.

Moreover, these results are in conflict with those from CDMS \cite{CDMS}, XENON10 \cite{XENON10}, ZEPLIN-II \cite{ZEPLIN-II} and ZEPLIN-III \cite{ZEPLIN-III}, which, on the contrary, found no evidence of annual modulation.

Directional detection of dark matter was proposed in \cite{Spergel:1988}, where it was observed that a strong forward-backward asymmetry is expected in the case of an isothermal spherical Galactic halo. 
Because of this, directional detection, even if in the case of poor angular resolution, might also provide information on the apparent WIMP wind direction.
Several experiments, including DRIFT \cite{DRIFT}, NEWAGE \cite{NEWAGE}, MIMAC \cite{MIMAC} and DMTPC \cite{DMTPC}, are focused on this goal.
Each of them is based on specific detection techniques.

DRIFT, NEWAGE, MIMAC and DMTPC are DM directional detectors using as target materials  low pressure gas mixtures. Nuclear recoils are identified by  measuring the induced ionization  with a time projection chamber (TPC)  technique.
 
DMTPC set a $90\%$ CL upper limit on the spin-dependent WIMP-proton cross section of $2.0\cdot10^{-33}$ cm$^2$ for 115 GeV WIMPs \cite{DMTPC}, while DRIFT's exclusion curve for spin-dependent WIMP-proton interactions reaches 1.1 pb at 100 GeV \cite{DRIFT}. 

There are also other types of proposals. One is to use nuclear emulsions, consisting of AgBr crystals immersed in an organic gelatin, as both  the WIMP target and  the tracking detector by reconstructing the direction of the recoiled nucleus \cite{Alexandrov:2014gda}. Another one is to use DNA or RNA, so that the path of the recoiling nucleus can be tracked to nanometer accuracy \cite{Drukier:2012hj}.

In ~\cite{Capparelli:2014lua}, a  new type of directional detector was outlined. It consists of large arrays of CNTs which provide the target material for WIMP-nuclei collisions and an anisotropic sensitivity to the recoil direction, complemented by a 
readout technique to detect the emitted nuclei. Given the relatively low mass of the C nuclei, such detector would extend  the WIMP search to low masses  ($\approx 10$~GeV) if a suitable  ion detection threshold of few KeV might be attained.

\section{CNTs and the Machta-Zwanzig regime in infinite horizon billiards}
\label{s:elementare}
As a first approximation, we consider the CNT system to be a regular array of  $L \times L$ cylindrical obstacles and assume that the particles cannot penetrate them. 
Elastic collisions are also assumed.

\begin{figure}[htbp]
\centering
\begin{tabular}{cc}
\includegraphics[width=4.1cm]{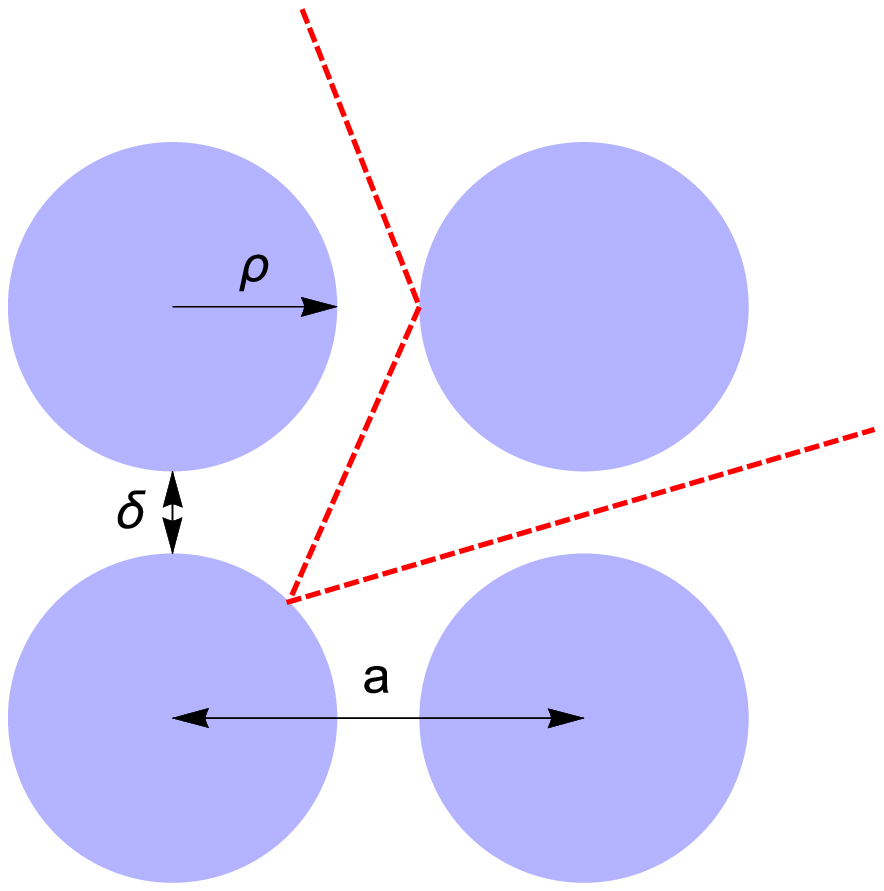} &\hspace{-0.5truecm} \includegraphics[width=4.1cm]{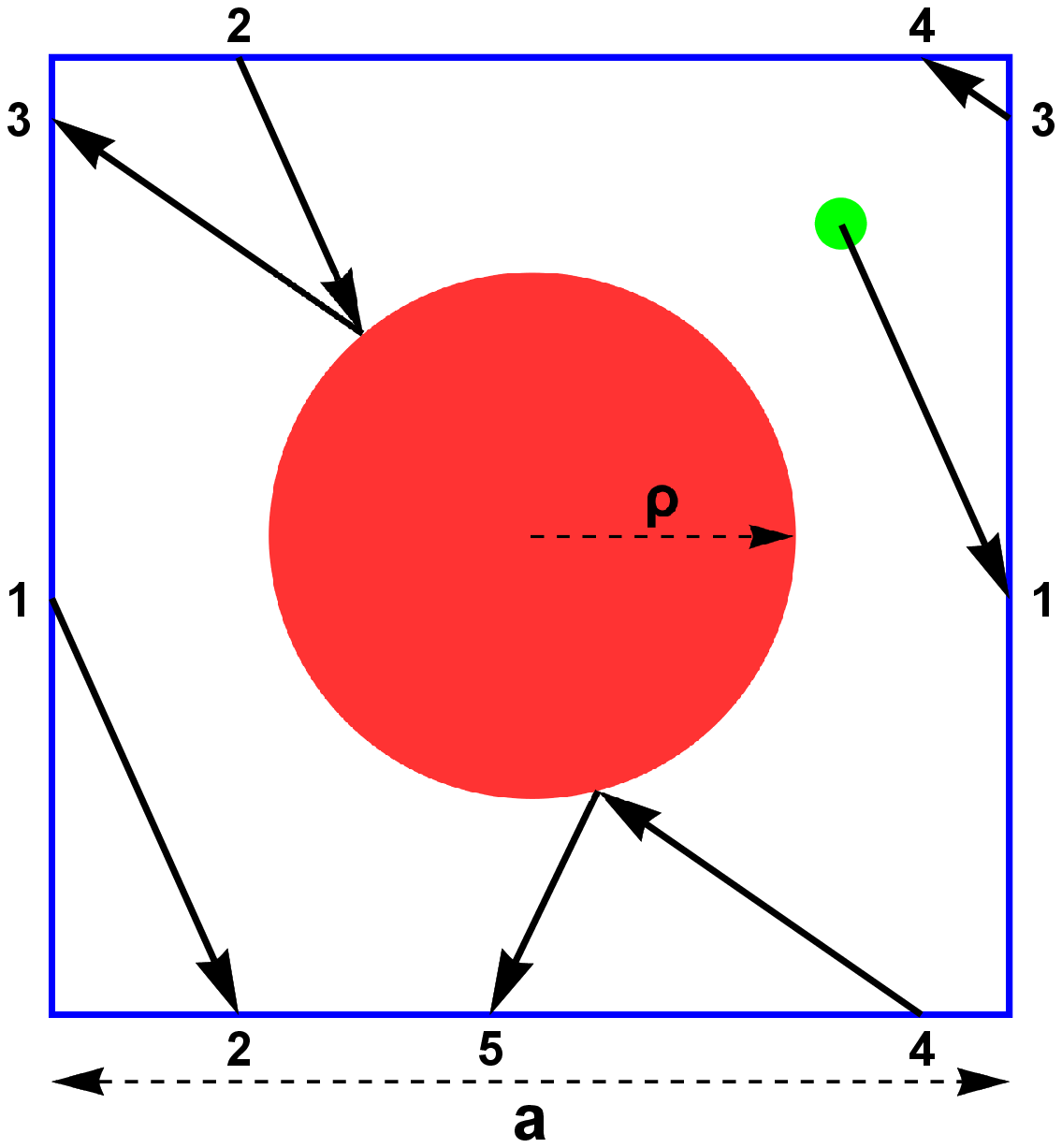}
\end{tabular}
\caption{Motion of the particle in the obstacle system, projected onto the 2D plane (left panel). Here, the circles represent the CNT sections, $a$ is the lattice spacing, $\rho$ the CNT radius and $\delta$ the inter-cylinder width. The ion motion in the lattice can also be treated as that of a particle within a primitive cell with periodic boundary conditions (right panel). The portion of trajectory shown in this picture starts at the green point. Positions connected by periodic boundary conditions are labeled by the same number.} 
\label{fig:primitive-cell}
\end{figure}

Let $h$ be the height of the cylindrical obstacles, $\rho$ the radius of their circular basis (see Fig. \ref{fig:primitive-cell}) and assume the cylinder axes are parallel to the $z$ direction (the  \textit{longitudinal} direction), in the $z>0$ half-space, their bases being on the $z=0$  plane (the \textit{transverse plane}). 
We call, respectively, the $z=0$ plane and its parallel plane at $z=h$ the \textit{bottom} and the \textit{top} ends of the system. Assume that the bottom bases centers of the $L^2$ cylinders form a square regular array on the $z=0$ plane. Let $a>2\rho$ be the distance between the centers of the bases of two neighboring cylinders and $\delta=a-2\rho$ the minimal width of the interstices (see Fig. \ref{fig:primitive-cell}, left panel).

Consider a particle entering the lattice through the bottom plane, thus with longitudinal velocity component $v_\parallel>0$, and let $v_\perp$ be the modulus of the projection of the velocity on the transverse plane.  

Assuming  uniformly distributed positions and transverse velocity directions as initial conditions at the bottom end,
we  estimate the typical time needed by a particle to exit the array from the top end.

The 2D motion of the system is equivalent to the motion of a particle inside an ``infinite-horizon'' periodic billiard table: a billiard in which there are corridors running straightforwardly towards the perimeter. The average time spent by a particle inside the billiard can be estimated  by using the so called Machta-Zwanzig approximation \cite{MZ,Cristadoro}, which holds in the limit $\delta\ll a$ (as for the finite-horizon case).  We assume therefore that the particle spends in each elementary cell (see Fig. \ref{fig:primitive-cell}, right panel) a random time exponentially distributed with mean 
\begin{equation}
	\label{e:ele000}
	\tau_{\rm R} = \frac{\pi(a^2-\pi\rho^2)}{4 v_\perp \delta}
\end{equation}
and then moves with equal probability $1/4$ to any of the four neighboring cells\footnote{For a derivation of Eq. (\ref{e:ele000}), see App. \ref{tauR}.}.
Hence, we can estimate the typical total time $\tau$ spent by a particle in the billiard by considering a two-dimensional homogeneous random walk on a square of $L \times L$ points in $\bb{Z}^2$.  The walker starts on the square from a uniformly distributed random point. We then evaluate the mean number of hops $n$ performed by the walker to reach the external boundary of the square. The typical time needed by the particle to exit the billiard is then given by 
\begin{equation}
	\label{e:ele010}
	\tau = n \tau_{\rm R} = n \mbox{ } \frac{\pi(a^2-\pi\rho^2)}{4 v_\perp \delta}
\end{equation}
We evaluate $n$ numerically by averaging the total number of hops performed by each particle in an ensemble of $10^6$ particles. By interpolating the results as a function of the side length $L$, we find $n=L^2/7$. 
The quadratic dependence on the side length of the square is quite obvious, since the random walk is symmetric, while the precise value of the numerical prefactor is not. 

In the following, we briefly discuss the algorithm we used in MC simulations for an infinite horizon billiard. The semi-analytical expression in  Eq. (\ref{e:ele010}), with $n=L^2/7$, is then compared to  MC numerical results.

\subsection{Simulated ions trajectories}
We consider a square primitive cell with side of length $a$ and periodic boundary conditions (Fig. \ref{fig:primitive-cell}, right panel).
The CNT transverse section is a centered circle in the cell. Each cell is labeled with the indices $(i,j)$, which define its position in the $L \times L$ lattice. 
A particle moves in the lattice and propagates through multiple reflections on obstacles. 
The starting position of the particle, projected onto the transverse plane, is determined by $(i_0,j_0)$, the initial $(x_0,y_0)$ coordinates within  the cell and the initial $v_{\perp,0}$ direction; these quantities are randomly extracted from uniform  distributions. We also assume that the particle trajectory starts outside the nanotube.

As a first step, we check if the ion trajectory intersects any obstacle in the cell: 1) If it does, we calculate the reflection angle and determine the new ion trajectory; 2) If it does not, we evaluate the ion position at the boundary of the cell, according to its direction. Then, we exploit the periodic boundary conditions to place the particle on the opposite side of the cell, with proper position and transverse velocity (Fig. \ref{fig:primitive-cell}, right panel). This procedure is repeated until a stop condition is fulfilled, that is either when the ion exits the lattice or it covers a length $d = h \mbox{ } \frac{v_\perp}{v_\parallel}$ in the transverse plane.

\subsection{Mean exit time}
\label{Simulations for a regular lattice}
Below, we show the MC results for the mean exit time, $\tau$, as a function of $L$ from a lattice with $\rho~=~5.4$ nm and $a=11$ nm. 
When the exit time $\tau$ from the transverse lattice is smaller than that from the top surface of the CNT array, we are losing a signal count in our ideal DM detector~\footnote{
CNT arrays are the target material for WIMP-nuclei collisions and must be complemented by a readout apparatus to detect those emitted ions which reach the top end of the system.}. 
  
\begin{figure}[htbp]
\centering
\includegraphics[width=5.5cm]{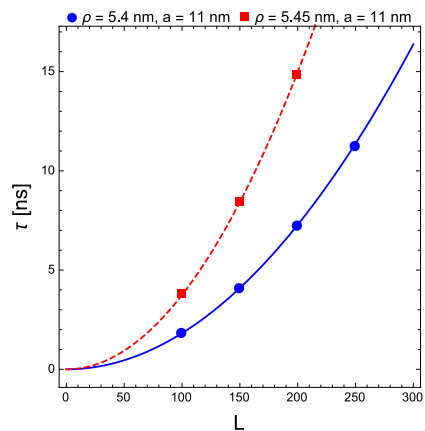}
\caption{Numerical values for the mean exit time from the transverse lattice ($\tau$), as functions of $L$, for $\rho = 5.4$ nm, $a = 11$ nm (circles) and $\rho =~ 5.45$~ nm, $a =~ 11$~ nm (squares) lattices. The point interpolation returns $p_0 = 9.13$ (continuos line) and $p_0 = 8.36$ (dashed line), respectively. The transverse velocity of the carbon ion is $v_\perp = \sqrt{2 E_\perp / M_C}$, where $E_\perp = 300$ eV is its transverse energy and $M_C$ its mass.} 
\label{fig:N-vari-fit}
\end{figure}

We evaluate $\tau$ for $L = 100, 150, 200, 250$ and we interpolate the results with $\tau (L) = \tau_{\rm R}\, \frac{L^2}{p_0}$, where $p_0$ is a free fit parameter.
The same is done for a $\rho=5.45$ nm and $a=11$ nm lattice.
The results are shown in Fig. \ref{fig:N-vari-fit}. 

\begin{figure}
\centering
\includegraphics[width=6cm]{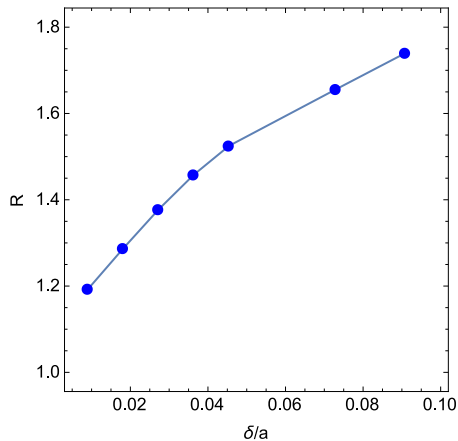}
\caption{Ratio $R$ between semi-analytical and numerical results for the mean exit time from the transverse lattice,  for different values of $\delta/a$.}
\label{fig:DifferentDeltaFit}
\end{figure}

Finally, in Fig. \ref{fig:DifferentDeltaFit}, we show the ratio $R$ of the semi-analytical and  numerical results for the average exit time from the transverse lattice, extracted for different values of $\delta/a$. As shown in Fig. \ref{fig:DifferentDeltaFit}, the Machta-Zwanzig  approximation becomes  effective once $\delta/a \rightarrow 0$.
 
\section{Lateral losses}
\begin{figure}[htbp]
\centering
\includegraphics[width=6cm]{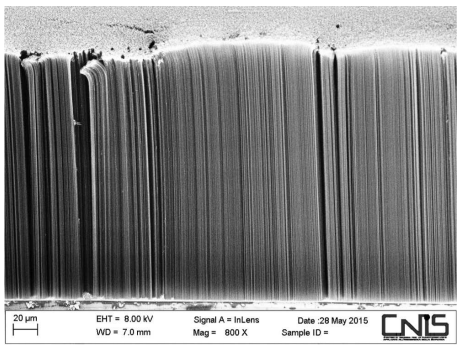}
\caption{Image of aligned CNTs on a silicon substrate obtained with a SEM. The sample has an inclination of $60^\circ$ with respect to the electron beam.} 
\label{fig:CNT-photo}
\end{figure}

In this Section, we address the question whether a particle reaches the top end before exiting the lattice through its sides.
We will consider arrays with larger lattice spacings, $a = 58$ nm, and $\rho = 5$ nm. These are the experimental  values  obtained with a scattering electron microscope (SEM) image analysis of  a  sample of aligned CNTs  (Fig. \ref{fig:CNT-photo}).

We then calculate the lateral losses through the sides in the case of a regular lattice and in the case of a random system of CNTs. In this second case, which reasonably provides a more realistic description of a physical CNT array~\footnote{In our study the CNT array is modeled in a simple but realistic way. Nevertheless, at this stage we neglected some other potential features of a physical CNT array, such as the presence of flaws (e.g. unused CNT growth precursors or not perfectly straight CNTs), which may reduce the CNT array channeling efficiency. 

}, the spacing between tubes is not constant. 
Specifically, we assume a Gaussian distribution for the displacements of the CNT center positions around the sites of a regular lattice.
The simulated carbon ion has an initial energy $E_\parallel ={\mathbf p}^2_\parallel/2M_C= 1$ keV and $E_\perp={\mathbf p}_\perp^2/2M_C = 300$ eV and it can either  exit the CNT lattice from the top, after covering the CNT length $h \simeq 300\, \mu$m, or from the lateral walls, i.e. exit from the transverse lattice, before the top is reached. In the latter case, a potential signal DM counting is lost.

We call $\eta$ the fraction of particles leaving from the external perimeter before reaching the top. We expect a $1/L$ behavior, which can be explained as follows. If $L$ is sufficiently large, we expect that most of the particles, which exit from the sides, originate within a thin frame close to the external perimeter.  The number of particles generated with random initial positions and momenta on the transverse lattice scales with $L^2$, whereas the number of particles generated close to the perimeter scales with $L$. Hence, we expect the $\eta \sim 1/L$ scaling.
\begin{figure}[htbp]
\centering
\includegraphics[width=6cm]{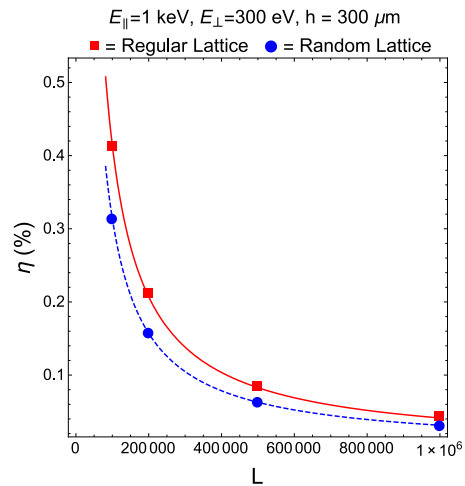}
\caption{Numerical results for $\eta$ on a transverse lattice with parameters $\rho = 5$ nm and $a = 58$ nm as a function of the lattice size $L$. $\eta$ stands for the fraction of particles leaving from the external lattice perimeter, before reaching the CNT top. We provide results for a regular (squares) and a random Gaussian lattice (circles), with zero mean $\mu$ and $\sigma = 10$ nm. The curves, $\eta_{\rm reg}(L)$ and $\eta_{\rm rand}(L)$, which fit the regular and random lattice values of $\eta$, are shown as continuous and dashed lines, respectively.}
\label{fig:exit-distance-plot}
\end{figure}

The results of our MC simulations for the regular and random lattices are shown in Fig. \ref{fig:exit-distance-plot} with  two curves superimposed, $\eta_{\rm reg}(L) = \frac{414}{L}$ and $\eta_{\rm rand}(L) = \frac{314}{L}$, which fit the data in the regular and random cases, respectively.
It is interesting to observe that  random lattices trap more particles than regular ones because infinite horizons are closed due to the random displacement of the obstacles. 

\begin{figure}[htbp]
\centering
\includegraphics[width=6cm]{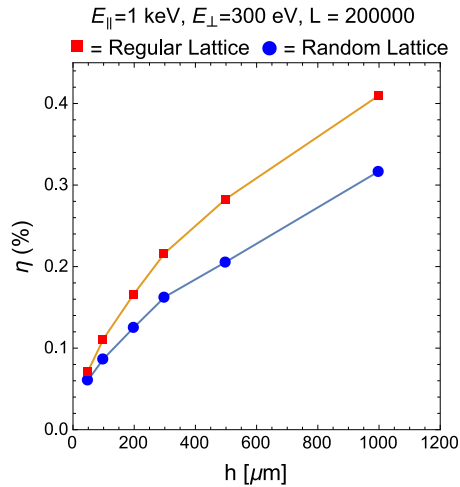} 
\caption{Numerical results of $\eta$ for a lattice with parameters $\rho = 5$ nm and $a = 58$ nm as a function of the CNT height $h$. $\eta$ stands for the fraction of particles leaving from the external lattice perimeter, before reaching the CNT top. We provide results for a regular (squares) and a random Gaussian lattice (circles), with null mean $\mu$ and $\sigma= 10$ nm.}
\label{fig:etaVSh-plot}
\end{figure}

In Fig. \ref{fig:etaVSh-plot}, we show another plot for $\eta$ as a function of the nanotube height, $h$, with $L = 2\times 10^5$. 
It is worthwhile noting that  lateral losses, if $L$ is sufficiently large, do not change dramatically as $h$ grows. Thus, relying on these results for $\eta$ as a function of $L$ and $h$, we can conclude that, if we consider a CNT array with $L$ of the order of $10^5$ or larger and a realistic value of the height, $h\sim$ 300 $\mu$m, the amount of lateral losses can be kept negligibly small.
These are explanatory results, taken in the limit of small ion longitudinal energy. An estimation of the lateral losses for ions whose initial conditions are produced by the scattering of a galactic halo WIMP on the target nucleus, as computed in \cite{Capparelli:2014lua}, are discussed in Sec. \ref{Interactions} (see Table \ref{tab:final-results}).

\section{Interactions between ions and CNT walls}
\label{Interactions}
In the previous simulations, we discussed the special case of ions whose transverse energy is below the potential barrier of the CNTs, so that reflections from nanotubes occur in the transverse lattice. 
In the following, we will consider the general case of  ion initial conditions being produced by the scattering of a galactic halo WIMP on the target nucleus, as computed in \cite{Capparelli:2014lua}.
In order to run our Monte Carlo simulations, a specific value for the WIMP mass, $M_\chi$, must be chosen. Our choice is to consider WIMPs with a mass equal to that of the carbon-12 nucleus, because the largest energy transfer occurs when projectile and target masses are equal.

In  the case the channeling initial conditions are not met, ions do not get reflected at  nanotube's surfaces  and are also subject to energy losses \cite{Dedkov,ZBL,Nastasi,O'C-B} and angular deviations  \cite{Meyer:1971,Sigmund:1974}, due to the interaction with atoms when crossing nanotubes.
Thus, unlike Ref. \cite{Capparelli:2014lua}, the ion motion is not limited to the internal volume of a carbon nanotube, but 
ions can propagate within the entire CNT array with the possibility of traversing CNT surfaces.

Moreover, in addition to  {\it chiral} CNTs, we also include a $30\%$ fraction of {\it armchair} tubes, which is approximately the fraction of metallic nanotubes in a generic CNT sample \cite{Rao}.
The potential barrier of chiral CNTs is axially symmetric and equal to 390 eV, while the potential of armchair CNTs can be approximated as a square pulse shape \cite[Fig. 5]{Capparelli:2014lua}. For the sake of simplicity, the latter is written as 
\begin{equation}
	\label{eqn:U-armchair}
	U(r= \rho,\phi) = \left\{ \begin{array}{rl} 100 \mbox{ eV} & (n \mbox{ even}) \\ 
	390 \mbox{ eV} & (n \mbox{ odd}) \end{array} \right.
\end{equation}
Here, $\phi$ and $r$ are cylindrical coordinates, with $\frac{2 \pi n}{304} + \phi_0  < \phi < \frac{2 \pi (n+1)}{304} + \phi_0$, $\phi_0$ is a random phase, extracted from a uniform distribution in the $[0,2\pi]$ range, and 304/2 is the number of ``windows" in the tube wall, namely those portions of the CNT surface for which $U(r= \rho,\phi) = 100$ eV. 

\subsection{Energy loss}
\label{Energy loss}
Nuclear stopping prevails on the electronic one \cite{Nastasi} for very slow ions (with a kinetic energy of a few keV per nucleon). The latter can  thus be neglected, as a first approximation. Following Ref. \cite{Dedkov}, with some appropriate modifications,  we will discuss the formalism to calculate 
energy losses.

The position dependent nuclear stopping force is given by
\begin{equation}
	F_n(r) = -S_n(E) n_a(r)
\end{equation}
where $r$ is the radial coordinate in the plane transverse to the CNT axis and $n_a(r)$ the nuclear distribution, averaged over a cylindrical surface of radius $r$ coaxial with the CNT. It  can be approximated as
\begin{equation}
  n_a(r) = \frac{n_s}{\sqrt{2\pi} u_\perp} \mbox{ } e^{-\frac{(r-\rho)^2}{2u_\perp^2}}
\end{equation}
where $u_\perp = 8.6\cdot10^{-3}$ nm is the transverse thermal vibration amplitude at 300 K and $n_s = 38.0$ nm$^{-2}$ the atomic density on the CNT surface \cite{Capparelli:2014lua}. 
$S_n(E)$  depends on the interatomic potential chosen to model the ion-nucleus interaction.
\begin{figure}[htbp]
\centering
\includegraphics[width=3.5cm]{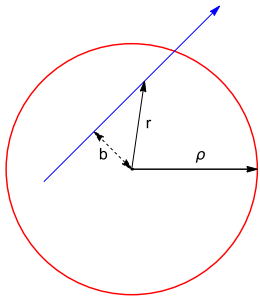}
\caption{The blue line represesents the trajectory of a particle crossing a CNT (red line), with radius $\rho$. $b$ is the impact parameter. Here, we neglect possible angular deviation effects.} 
\label{fig:DedkovBroken}
\end{figure}
Our choice is the ZBL (Ziegler, Biersack and Littmark) universal repulsive potential \cite{ZBL}, because it provides, in the energy range we consider here, results that are more precise than those of classical atomic models, like Thomas-Fermi or Bohr \cite{O'C-B}. In Ref. \cite{Nastasi}, the ZBL potential was interpolated  with  the formula
\begin{equation}
	S_n(E) = \frac{8.462\cdot10^{-15} Z_1 Z_2 M_1 S_n(\epsilon)}{(M_1+M_2) (Z_1^{0.23} + Z_2^{0.23})} \mbox{ eV}\cdot\mbox{cm}^2
\end{equation}
where $S_n(\epsilon) = \frac{0.5 \mbox{ log}(1+1.1383\epsilon)}{\epsilon+0.01321\epsilon^{0.21226}+0.19593\epsilon^{0.5}} $.
Here, the indices 1 and 2 refer to the projectile and target nuclei, respectively, and $\epsilon$ is the reduced energy, defined as
\begin{equation}
	\label{eqn:reduced-epsilon}
	\epsilon(E) = \frac{M_2 a E}{(M_1+M_2)Z_1 Z_2 e^2}
\end{equation}
where $ a = 0.885 \mbox{ } a_0 (Z_1^{2/3} + Z_2^{2/3})^{-1/2}$ is the interatomic screening radius in Lindhard's theory ($a_0$ is the Bohr radius), $E$ the ion kinetic energy, $M_i$ and $Z_i e$ ($i= 1,2$) the masses and charges of the projectile and target nuclei.

We call $b$ the impact parameter with respect to the center of the nanotube (Fig. \ref{fig:DedkovBroken}).
When the ion crosses the CNT wall, it loses an amount of energy
\begin{equation}
	\label{eqn:DeltaT}
	\Delta E(b;E) = \int _0^\infty F_n\left(\sqrt{x^2+b^2}\right) dx
\end{equation}
where $x = \sqrt{r^2 - b^2}$.  
It is then convenient to average $\Delta E(b;E)$ over the impact parameter distribution of an homogeneous beam, $\frac{dP}{db} = \frac{4 \sqrt{\rho^2 - b^2}}{\pi \rho^2}$.
In this way, we obtain a parametrization for the average energy loss in a CNT crossing depending on ion kinetic energy only.

\subsection{Angular deviation effects}
\label{Angular deviation effects}   
When the ion crosses a nanotube surface, it is also subject to angular deviations due to Coulomb scattering with 
nuclei. 
We consider two possible regimes, depending on the angle of the incident particle with respect to the CNT surface: 1. single scattering 2. multiple scattering. A rough way to distinguish between them is discussed in the following.

\begin{figure}[htbp]
\centering
\includegraphics[width=5cm]{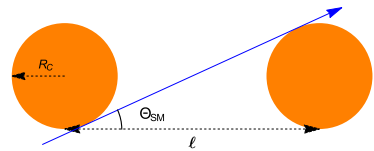}
\caption{2D representation of the procedure used to estimate $\Theta_{SM}$. The blue line is the limit ion trajectory, which distinguishes between single and multiple scattering regimes. The orange circles are the screened Coulomb fields (of radius $R_C$) of two carbon nuclei  in the graphene lattice, belonging to the same unit cell. $\ell = 0.14$ nm is the carbon-carbon bond length in graphene.} 
\label{fig:thetaMS}
\end{figure}

We introduce here a limit angle $\Theta_{SM}$. If the incident angle between the ion direction and the CNT surface is larger than $\Theta_{SM}$, the ion is subject to single (Rutherford) scattering, otherwise undergoes multiple scattering. The incident angle is equal to $\Theta_{SM}$ when the ion trajectory, as it crosses the nanotube surface, is tangent to the screened Coulomb fields of two carbon atoms belonging to the same unit cell. The Coulomb fields are described as spheres, with radius $R_{\rm C} = a_0 \, Z^{-1/3}$. 
$\Theta_{SM}$ can then be calculated as (see Fig. \ref{fig:thetaMS})
\begin{equation}
	\Theta_{SM} = \arcsin \left(2R_{\rm C}/\ell\right)
\end{equation}
where $\ell = 0.14$ nm is the carbon-carbon bond length in graphene.

\subsubsection{Single scattering regime}
If the particle enters the CNT with an incident angle larger than $\Theta_{SM}$, we can calculate the ion-nucleus scattering angle using Rutherford theory.
In particular, the scattering angle, $\theta_{\rm s}$, is extracted from the distribution $\sim$  $\frac{\mbox{cos } \theta_{\rm s}}{\mbox{sin}^3 \theta_{\rm s}}$.

The probability of carbon-ion--nucleus scattering can be estimated as the ratio between the fraction of surface of the graphene lattice unit cell ``occupied'' by the screened nuclear Coulomb field and the total area of the unit cell. With an effective shielding radius of the order of $a_0\, Z^{-1/3}$, the calculated probability is $10\%$, approximately.

\subsubsection{Multiple scattering regime}
When a scattered carbon ion moves grazing a CNT wall (a graphene layer) with a direction almost parallel to its axis, multiple scattering might occur.

The most appropriate treatment to estimate angular deviations of low energy ions undergoing  $N \lesssim 20$ scatterings, is described in~\cite{Meyer:1971,Sigmund:1974}.
The starting point is the differential scattering cross-section by Lindhard, Nielsen and Scharff (LNS) \cite{LNS:1968}, which describes the dynamics of a projectile incident upon a target atom on the basis of classical scattering theory and the Thomas-Fermi model of  atoms. 
In the treatment  in~\cite{Meyer:1971,Sigmund:1974} further assumptions are considered: a) atoms are described as spheres with radius $r_0$; b) binary collision events have azimuthal symmetry; c) there is negligible energy loss in any single collision; d) the scattering angles are small.
In this way a simplified expression of the angular distribution of the projectile after multiple scattering can be obtained in terms of reduced variables~\cite{Meyer:1971}. These are the reduced energy, $\epsilon$ [see Eq. (\ref{eqn:reduced-epsilon})], the reduced total scattering angle, $\tilde \theta_{\rm s} = \frac{\epsilon}{2} \mbox{ } \frac{M_1+M_2}{M_2} \mbox{ } \theta_{\rm s}$, and the reduced thickness, $\chi = \frac{a^2 N}{r_0^2}$, given in dimensionless units.
Under these assumptions, the angular distribution of the projectile after multiple scattering can be written as
\begin{equation}
	\begin{array}{l}
	F(\tilde \theta_{\rm s}) = \frac{\epsilon^2}{8 \pi} \left(\frac{M_1+M_2}{M_2}\right)^2 \left[ f_1 (\chi,\tilde \theta_{\rm s}) - \frac{a^2}{r_0^2} 
	f_2 (\chi,\tilde \theta_{\rm s}) \right]
	\end{array}
\end{equation}
where $f_1(\chi,\tilde \theta_{\rm s})$ and $f_2(\chi,\tilde \theta_{\rm s})$ are functions, calculated by numerical integration and tabulated in Ref. \cite{Meyer:1971}.

\subsection{Monte Carlo estimates of the CNT array channeling efficiency}
\label{Monte Carlo estimates of the CNT array efficiency}
We consider a random lattice, with parameters $a$ = 58 nm, $\rho$ = 5 nm, $\mu = 0$, $\sigma = 10$ nm, size $L = 10^5$ and nanotube height of 300 $\mu$m. The open end of the CNT array is directed along the WIMP wind direction, which is opposite to the Sun's direction in the Galactic frame (leading towards the Cygnus constellation). 
The particle trajectory is considered to start outside the nanotube internal volume at some given distance  from the bottom of the CNT array basis extracted from a uniform random distribution. 
We define different event categories. 1) {\it Top events}, when the ions  exit from the top end of the array, with energy above a detection threshold on the ion kinetic energy of  $E_{\rm th} = 1$ keV\footnote{At this stage, this  threshold of the readout apparatus  is ideal and further experimental tests are needed.
}: these are the ions that can eventually be detected; 2) {\it Bottom events}, when the ions reach the CNT substrate;  3)~{\it Dead events}, when the ions reach energies  below $E_{\rm th}$ (because of energy loss) while they are still propagating within the array; 4) {\it Side events}, when the ions exit from the lateral sides of the array.  
For all of them, the initial kinematic conditions are taken from  distributions expected for WIMP-nuclei collisions. 

We  define $\theta$ to be the angle between the ion trajectory and the Sun's direction.

\begin{table}[htbp]
\caption{Results of our simulation. We consider a total number of  $10^5$  ions trajectories and report the fraction of possible Top, Bottom, Side and Dead events, as explained in the text.}
\label{tab:final-results}
\begin{tabular}{cccc} 
\hline\noalign{\smallskip}
Top     & Bottom & Side & Dead  \\
\noalign{\smallskip}\hline\noalign{\smallskip} 
$(30.8\pm0.1) \%$ & $(1.79\pm0.04)\%$ & $(0.45\pm0.02)\%$  & $(67.0\pm0.1)\%$ \\
\noalign{\smallskip}\hline
\end{tabular}
\end{table}

\begin{figure}[htbp]
\centering
\includegraphics[width=5.5cm]{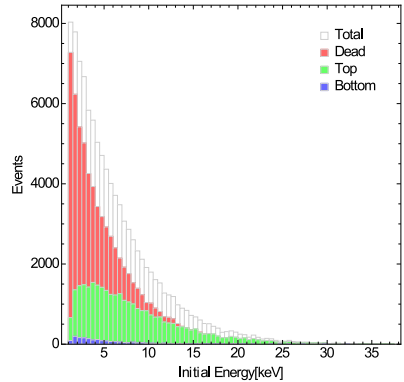}
\caption{Distribution of the initial total kinetic energy $E=E_\parallel+E_\perp$. We also show the distributions of top, dead and bottom events as a function of the initial conditions.} 
\label{fig:initial-distr-E}
\end{figure}

\subsubsection{Number of top events}
\begin{figure}[htbp]
\centering
\includegraphics[width=5.5cm]{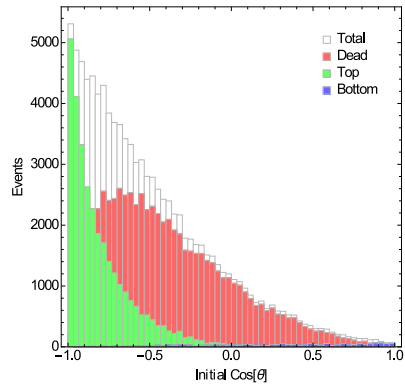}
\caption{Distribution of the initial ion directions. We also show the distributions of top, dead and bottom events as a function of the initial conditions.} 
\label{fig:initial-distr-Cos}
\end{figure}

\begin{figure}
\centering
\includegraphics[width=5.5cm]{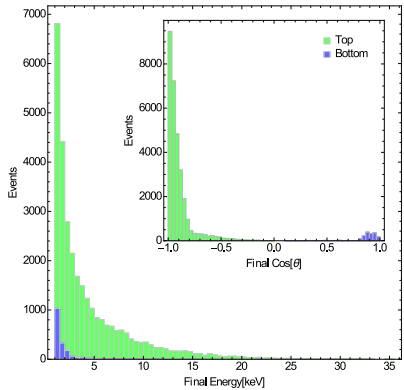}
\caption{Histograms of final ion kinetic energies and final directions for top and bottom events.} 
\label{fig:final-distr-E-Cos}
\end{figure}

A MC simulation including all the effects described above is performed to  study the CNT array channeling efficiency.  The latter is defined as the ratio between top events and the total number of scattering events.
The results are shown in Figs. \ref{fig:initial-distr-E}--\ref{fig:final-distr-E-Cos} and Table \ref{tab:final-results}. The estimated channeling efficiency of the CNT array is 31\%.

When the initial ion energy distribution is peaked at low values of $E$, as in Fig.~\ref{fig:initial-distr-E}, most of the particles do not succeed to exit the array, as their energy sets below $E_{\rm th}$ before they can leave the lattice. 
On the other hand, the target channeling efficiency heavily depends on the orientation of the CNT, as expected; in particular, since the recoil distribution is peaked at initial angles $\theta \approx \pi$, Fig. \ref{fig:initial-distr-Cos}, the number of top events is much larger than that of bottom events.

This is an important confirmation of the hypothesis of an anisotropic response of the CNT array to the WIMP wind described in  \cite{Capparelli:2014lua}.
Moreover, almost $1/4$ of top events leave the array because of {\it secondary}  channeling effects, which can change  the initial particle direction and make it satisfy  the  channeling conditions in the CNT interior.

In Fig. \ref{fig:final-distr-E-Cos}, we show the final ion kinetic energies and directions as the particles reach the top or the bottom of the CNT array. The ions might roughly lose on average 4 keV during their path towards the top of the array.
As expected, top events take place when ions have an initial kinetic energy higher than the mean initial one; moreover, we observe that their mean final angle is closer to $\theta = \pi$ than the mean initial one, due to secondary channeling effects. 

\subsubsection{Target mass}
After providing an estimation of the target channeling efficiency for a given set of lattice parameters, we want to see how it varies with the lattice spacing at fixed CNT radius, $\rho = 5$ nm. 
This can give important indications to optimize the cost and size of the target. For example, if distances between tubes are small, we can arrange more CNTs on  a given surface. 

For small values of $a$, particles undergo many collisions with CNTs. Therefore, their mean energy loss is larger and the number of dead events increases with respect to the large corridor case.
However, a relatively strong decrease of the lattice spacing does not heavily reduce the number of top events. Specifically, the channeling efficiencies of a lattice with $a = 10$ nm and one with $a = 70$ nm are $27\%$ and $31 \%$, respectively.
\begin{table}[htbp]
\caption{Total CNT mass in a large-scale target, as a function of the lattice spacing $a$. Nanotube height is $h$ = 300 $\mu$m and radius is $\rho = 5$ nm.} 
\label{tab:detector-mass}  
\begin{tabular}{cc} 
\hline\noalign{\smallskip}
$a$ [nm] & \hspace{0.2 cm} Target mass [kg] \\ 
\noalign{\smallskip}\hline\noalign{\smallskip}
11           & 11.8  \\
30           & 1.6  \\
45           & 0.7  \\
58           & 0.4  \\
\noalign{\smallskip}\hline
\end{tabular}
\end{table}

Given this, we want to find a value for the lattice spacing, which may provide a suitable target mass and a good channeling efficiency. We assume that the total surface of the target is 100 m$^2$. The target is made up of 100 panels. Each of them has a surface of 1 m$^2$ and consists of a double ``brush", composed of a large number of adjacent CNT arrays. 
Including the active detection volume of the readout apparatus, each panel might have a thickness of $\approx 10$ mm.
The target plus detector volume is then roughly $1 \times 1 \times 1$ m$^3$. We assume that CNT array units are closely packed with each other, so that the number of CNTs in a panel is $\approx 2 \mbox{ } \frac{\mbox{ m}^2}{a^2}$.
Considering a CNT surface density of $1/1315$ g$\cdot$m$^{-2}$ and a nanotube height of 300 $\mu$m, we compute the total CNT mass of the target for different values of $a$. The results are reported in Table \ref{tab:detector-mass}.
In order to have about 1 kg of CNT mass, we should consider arrays with distances between nanotubes less than 30 nm. These values for $a$ correspond to an array channeling efficiency of $27\%$.

\section{Results and conclusions}
We studied the dynamics of low energy ions in a carbon nanotube lattice with Monte Carlo simulations. 
Our results were used to provide an evaluation of the channeling efficiency of a target, composed of a large number of adjacent CNT arrays. 
The system, consisting of the CNT lattice target for WIMP particles plus the readout apparatus to detect the ions emitted by the lattice at CNTs' top end, represents our WIMP directional detector \cite{Capparelli:2014lua}.

At the present initial stage of this research, we only focus on the channeling potential of our target, as a function of parameters like the lattice spacing and nanotubes height/diameter. 
The target is considered to be background-free, even though it may contain a certain amount of radioactive contaminants. A detailed discussion of  background sources and an exhaustive description of the readout apparatus needs an experimental effort that is underway.

Collisions with WIMPs might struck carbon ions out of their lattice positions and inject them in the array with initial kinematical conditions depending on the relative orientation of the CNT axes with respect to the apparent WIMP wind from the Cygnus. 
The smaller the angle between these two directions, the higher the probability of channeling within the single nanotubes or from the interstices among them. The ions whose trajectories exit from the surface with open end nanotubes (on the opposite side CNTs terminations are closed by a substrate as in Fig.~\ref{fig:CNT-photo}), might be detected by the readout apparatus and constitute  the directional DM signal.  

The array efficiency depends on several technical specifications, as the length of the CNT array basis in terms of number of nanotubes and the lattice spacing.
Ions with right initial conditions, 
which anyway get lost from the lateral sides of the array, are weakening the signal. 
It is thus worthwhile to estimate the lateral ion losses as a function of the size of the array.
We showed that lateral ion losses can be kept sufficiently low if we consider CNT arrays with a number of nanotubes $\gtrsim 10^5\times 10^5$, within a large range of nanotube heights. As can be seen from the results of Table~\ref{tab:final-results}, calculated with the initial kinematic conditions from the ion recoil distribution (see \cite{Capparelli:2014lua} and Figs.~\ref{fig:initial-distr-E}, \ref{fig:initial-distr-Cos}), the expected lateral losses are of the order of $0.5\%$, approximately.  

Therefore, commercially available CNT arrays, with a lateral side of $10^6 - 10^7$ nanotubes, are substantially unaffected by this problem.
The channeling efficiency should also depend on the lattice spacing, $a$, between nanotubes -- in a weak way as we found. Because of this, we can consider a small lattice spacing, $a < 30$ nm, in order to pack a suitable target mass, of the order of 1 kg, on a given surface (100 m$^2$, approximately) preserving, at the same time, the array channeling efficiency.

An important outcome of the present study and Ref. \cite{Capparelli:2014lua} is that a CNT array may cooperate to effectively provide the required anisotropic response for the directional detection of WIMPs.
In fact, the number of calculated top events is much larger than that of bottom events, which corresponds to the number of top events when the CNT brush is reversed with respect to the direction of the WIMP wind (Table \ref{tab:final-results}).
The double CNT brush depicted in Ref. \cite{Capparelli:2014lua} might be able to help to  discriminate the signal from the background. This would  necessarily involve, for example, an assessment  of the radiopurity of the CNTs, which is out of the scope of the present paper. 

In \cite{Capparelli:2014lua} the analysis was focused on the calculation of ion channeling within the fundamental constituent of the lattice: a single nanotube. 

It was shown that it is difficult to match the right channeling conditions inside a single CNT immediately after a WIMP collision. 
However the CNT array as a whole -- including interstices among nanotubes --  can enlarge the acceptance of the target cooperating at driving the trajectories of not perfectly collimated ions  towards the open exit. Indeed, we showed here that those ions which cannot be channeled in a single nanotube, but have initial kinematic conditions close to the channeling ones,  have further chances to be channeled by the array, including the  interstitial space between CNTs.

In Ref. \cite{Capparelli:2014lua}, the calculated fraction of ions channeled in the nanotubes, immediately after a WIMP scattering event, is $0.4\%$; see \cite[Fig. 7]{Capparelli:2014lua}. The remaining large majority, which is the subject of the present study, consists of those particles scattered outside  the single nanotube and/or not satisfying the very  strict channeling conditions within a single CNT.
As shown in Table \ref{tab:final-results}, a non-negligible fraction of these particles 
may be guided towards the top end of the system by the interstitial spaces between CNTs. 
Therefore, by combining the previous results of Ref. \cite{Capparelli:2014lua} and those of the present paper, we estimate that the average fraction of particles  reaching the top end of the system  grows from $0.4\%$ \cite{Capparelli:2014lua} up to  $\approx 30\%$.  This represents  the  channeling efficiency  for ions  of our CNT target after a WIMP scattering event. 
     
To quantify the channeling potentiality of our CNT lattice as a whole with respect to the single nanotube case, we can also introduce an acceptance angle, $\theta_{\rm a}$.
Specifically, the acceptance angle is calculated in such a way that the integral over the ion recoil distribution up to $\theta_{\rm a}$ is equal to the fraction of channeled top events, namely those events which reach the top end satisfying channeling conditions.
The computed value for the CNT array is $\theta_{\rm a} = 35^\circ$, while that for the single nanotube, calculated by using the findings of Ref. \cite{Capparelli:2014lua}, is $\theta_{\rm a} = 4^\circ$.
Thus, an increase in the channeling efficiency of the lattice as a whole, with respect to the single nanotube, can be translated in terms of a larger target acceptance.

Such an improvement has obvious consequences on the construction specifications of a Dark Matter detector, especially in terms of target mass.  Testing experimentally the single carbon nanotube array unit, with electron or neutron probes for example, will finally tell to which extent the approximation used in our calculations are realistic. 

\begin{acknowledgements}
We thank M. Diemoz for supporting this research activity and  M.G. Betti and C. Mariani 
for giving access to the CNIS facility at Sapienza Univ. Roma. We acknowledge former collaboration on these topics with L.M. Capparelli and D. Mazzilli. 
G.C. acknowledges partial support from ERC  Ideas Consolidator Grant CRYSBEAM G.A. n.615089.
E.N.M.C. thanks M. Lenci for many helpful discussions. 

\end{acknowledgements}

\begin{appendix}

\section{Residence time in Machta-Zwanzig regime}
\label{tauR}
\begin{figure}
\centering
\includegraphics[width=8cm]{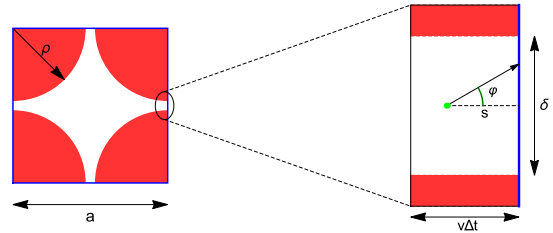}
\caption{Square elementary cell (left panel). The portion of the cell area the particle can escape from in a time smaller than $\Delta t$ is zoomed in on the right. The particle is a distance $s$ from the slit and has direction $\varphi$. It can leave the cell in a time less than $\Delta t$ if $s < v \Delta t \mbox{ cos } \varphi$.} 
\label{fig:MZ}
\end{figure}

Following Ref. \cite{MZ},  we calculate the residence time, $\tau_{\rm R}$, of a particle in an infinite-horizon billiard with an elementary square cell. See Fig. \ref{fig:MZ} (left panel).

For $\delta \rightarrow 0$, the particle motion in the elementary square cell can be regarded as ergodic. 
Thus, the probability that a particle leaves the cell in a time smaller than $\Delta t$ is given by the ratio between the phase space available to exit, which we call $\Omega$, and the total phase space, $\Omega_0$.
The total volume of the phase space of the cell is $\Omega_0=2 \pi vA$, with $A = a^2 - \pi \rho^2$. The calculation of the phase space available to exit the cell is more complex and is discussed in the following.

We refer to Fig. \ref{fig:MZ}. We assume that, due to the smallness of $v \Delta t$, the obstacle curvature can be ignored. A particle, which is at distance $s$ away from the border, leaves the cell in a time smaller than $\Delta t$ if $s < v \Delta t$ and $\varphi <\arccos \left(\frac{s}{v \Delta t}\right)$.
Thus, the portion of velocity space for the particle to exit the cell has an area $2 v \arccos \left(\frac{s}{v \Delta t}\right)$. 
We can now integrate over the rectangle $\delta \times v \Delta t$ of Fig. \ref{fig:MZ} (right panel) and obtain
\begin{equation}
	\Omega=4 \delta \int_0^{v \Delta t} ds \mbox{ } 2 v \arccos \left(\frac{s}{v \Delta t}\right) = 8 \delta v^2 \Delta t
\end{equation} 
where we also have a factor of four because there are four possible exit paths.
The particle exits the cell when $\Omega / \Omega_0 $ $= \frac{4 \delta v \Delta t}{\pi A} = 1$. 
The average residence time in the cell is thus
\begin{equation}
	\tau_{\rm R} = \Delta t = \frac{\pi A}{4\delta v} = \frac{\pi (a^2 - \pi \rho^2)}{4\delta v}
\end{equation}	 

\end{appendix}


\begin{thebibliography}{}

\bibitem{Capparelli:2014lua} 
  L.~M.~Capparelli, G.~Cavoto, D.~Mazzilli and A.~D.~Polosa,
  Phys.\ Dark Univ.\  {\bf 9-10}, 24 (2015);
  L.~M.~Capparelli, G.~Cavoto, D.~Mazzilli and A.~D.~Polosa, 
  Corrigendum to Phys.\ Dark Univ.\  {\bf 9-10}, 24 (2015).

\bibitem{Lindhard-channeling}
  J. Lindhard, 
  Kongel. Dan. Vidensk. Selsk., Mat.-Fys. Medd. {\bf 34}, No. 14 (1965).
  
\bibitem{Artru}  
  X. Artru, S. P. Fomin, N. F. Shulga, K. A. Ispirian and N. K. Zhevago,  
  Phys. Rept. {\bf 412}, 89 (2005).   
  
\bibitem{Dedkov}  
  G. V. Dedkov,
  Surf. Coat. Technol. {\bf 158-159}, 75 (2002).  

\bibitem{MZ}
 J.\ Machta and R.\ Zwanzig,
 Phys.\ Rev.\ Lett. \textbf{50}, 1959 (1983).

\bibitem{Cristadoro}
  G.\ Cristadoro, T.\ Gilbert, M.\ Lenci, and D.P. Sanders, 
  Phys.\ Rev.\ E \textbf{90}, 022106 (2014);
  \textbf{90}, 050102(R) (2014).
 
\bibitem{ZBL}
  J. F. Ziegler and J. P. Biersack,
  {\it The stopping and range of ions in matter}, Springer, 1985.  
  
\bibitem{O'C-B}
  D. J. O'Connor and J. P. Biersack, 
  Nucl. Instrum. Meth. Phys. Res. B {\bf15}, 14 (1986).
  
\bibitem{Nastasi}
  M. Nastasi, J. Mayer and J. K. Hirvonen, 
  {\it Ion-solid interactions: fundamentals and applications}, 
  Cambridge University Press, 1996.  
  
\bibitem{Meyer:1971}  
  L. Meyer, 
  Phys. Status Solidi B {\bf 44}, 253 (1971).
  
\bibitem{Sigmund:1974}
  P. Sigmund and K. B. Winterbon, 
  Nucl. Instrum. Meth. {\bf 119}, 541 (1974). 
  
\bibitem{Bernabei:2015xba} 
  R.~Bernabei {\it et al.} [DAMA Collaboration],
  Int. J. Mod. Phys. D {\bf 13}, 2127 (2004);
  Eur. Phys. J. C {\bf 56}, 333 (2008);
  Phys.\ Part.\ Nucl.\  {\bf 46}, no. 2, 138 (2015).   
  
\bibitem{CRESST} 
  G. Angloher {\it et al.} [CRESST Collaboration],
  Astropart. Phys. {\bf 23}, 325 (2005);
  {\bf 31}, 270 (2009).    
  
\bibitem{CoGeNT}  
	C. E. Aalseth {\it et al.} [CoGeNT Collaboration],
	Phys. Rev. Lett. {\bf 106}, 131301 (2011);
	Phys. Rev. Lett. {\bf 107}, 141301 (2011).
  
\bibitem{CDMS}
  D. S. Akerib {\it et al.} [CDMS Collaboration],
  Phys. Rev. D {\bf 73}, 011102 (2006);
  Phys. Rev. Lett. {\bf 96}, 011302 (2006);
  Z. Ahmed {\it et al.} [CDMS Collaboration],
  Phys. Rev. Lett. {\bf 102}, 011301 (2009).

\bibitem{XENON10}
  J. Angle {\it et al.} [XENON10 Collaboration],
  Phys. Rev. Lett. {\bf 100}, 021303 (2008);
  Phys. Rev. D {\bf 80}, 115005 (2009).

\bibitem{ZEPLIN-II} 
  G. J. Alner {\it et al.} [ZEPLIN-II Collaboration],
  Astropart. Phys. {\bf 28}, 287 (2007).

\bibitem{ZEPLIN-III}
  V. N. Lebedenko {\it et al.} [ZEPLIN-III Collaboration],
  Phys. Rev. D {\bf 80}, 052010 (2009).
  
\bibitem{Spergel:1988}
  D. N. Spergel, 
  Phys. Rev. D {\bf 37}, 1353 (1988).  
  
\bibitem{DRIFT}  
  G. J. Alner {\it et al.},
  Nucl. Instrum. Meth. A {\bf 555}, 173 (2005);
  S. Burgos {\it et al.}, Astropart. Phys. {\bf 31}, 261 (2009);
  J.~B.~R.~Battat {\it et al.} [DRIFT Collaboration],
  Phys.\ Dark Univ.\  {\bf 9-10}, 1 (2014).
     
\bibitem{NEWAGE}
  K. Miuchi {\it et al.} [NEWAGE Collaboration],
  Phys. Lett. B {\bf 654}, 58 (2007);
  {\bf 686}, 11 (2010).  
  
\bibitem{MIMAC}
  D. Santos {\it et al.} [MIMAC Collaboration],
  J. Phys. Conf. Ser. {\bf 65}, 012012 (2007);
  {\bf 460}, 012007 (2013);
  Q.~Riffard {\it et al.} [MIMAC Collaboration],
  arXiv:1504.05865.
  
\bibitem{DMTPC}
  S.~Ahlen {\it et al.} [DMTPC Collaboration],
  Int.\ J.\ Mod.\ Phys.\ A {\bf 25}, 1 (2010);
  Phys. Lett. B {\bf 695}, 124 (2011);
  J. P. Lopez {\it et al.} [DMTPC Collaboration],
  Nucl. Instr. Meth. A {\bf 696}, 121 (2012).  
  
\bibitem{Alexandrov:2014gda} 
  A.~Alexandrov {\it et al.},
  JINST {\bf 9}, no. 12, C12053 (2014).  
  
\bibitem{Drukier:2012hj} 
  A.~Drukier, K.~Freese, A.~Lopez, D.~Spergel, C.~Cantor, G.~Church and T.~Sano,
  arXiv:1206.6809 [astro-ph.IM].  
  
\bibitem{Freese:2012xd} 
  K.~Freese, M.~Lisanti and C.~Savage,
  Rev.\ Mod.\ Phys.\  {\bf 85}, 1561 (2013).  
    
\bibitem{Rao}
  C. N. R. Rao and A. Govindaraj, 
  {\it Nanotubes and nanowires},  Royal Society of Chemistry, 2011.    

  
  
  
\bibitem{LNS:1968}
  J. Lindhard, V. Nielsen and M. Scharf, 
  K. Danske Vidensk. Selsk. Mat.-Fys. Medd. {\bf36}, No. 10 (1968). 
  

  
  
  
  

  
  
  


\end{thebibliography}
\end{document}